\documentclass[10pt]{iopart}

\usepackage{iopams}
\usepackage{cases}
\usepackage{graphicx}
\usepackage[USenglish]{babel}
\usepackage{xcolor}

\usepackage{hyperref}
\usepackage[normalem]{ulem}
\usepackage{etoolbox}
\patchcmd{\subequations}{\alph{equation}}{\textit{\alph{equation}}}{}{}
\usepackage[capitalize]{cleveref}
\Crefname{figure}{Fig.}{Figs.}
\Crefname{equation}{Eq.}{Eqs.}
\Crefname{section}{Sec.}{Secs.}
\Crefname{subsection}{Sect.}{Sects.}

\newcommand{\qty}[2]{{#1}\;{#2}}
\newcommand{\unit}[1]{{#1}}
\newcommand{\text}[1]{\mbox{\textrm{#1}}}
\newcommand{\giga}{\text{G}}
\newcommand{\hertz}{\text{Hz}}
\newcommand{\kelvin}{\text{K}}
\newcommand{\tesla}{\text{T}}
\newcommand{\milli}{\text{m}}
\newcommand{\nano}{\text{n}}
\newcommand{\metre}{\text{m}}
\newcommand{\meter}{\text{m}}

\newcommand{\centi}{\text{c}}

\newcommand{\m}{\text{m}}
\newcommand{\degreeCelsius}{$^\circ\mathrm{C}$}
\newcommand{\electronvolt}{\text{eV}}

\newcommand{\mathmodeifneeded}[1]{%
  \ifmmode
    {#1} 
  \else
    ${#1}$ 
  \fi
}
\newcommand{\ohm}{\mathmodeifneeded{\mathrm{\Omega}}}%
\newcommand{\squared}{\mathmodeifneeded{^{2}}}%

\usepackage{isomath}

\let\vec\vectorsym

\newcommand{\Zf}{Z_{film}}

\newcommand{\rhovm}{\rho_{vm}}

\newcommand{\tauQP}{\tau_{core}}

\newcommand{\rhoff}{\rho_{\it ff}}
\newcommand{\Borb}{B_{c2}^{orb}}
\newcommand{\Bpau}{B_{c2}^{Pauli}}

\newcommand{\omegatau}[1]{\langle \omega_c \tauQP \rangle_{\mathrm{#1}}}

\usepackage{xspace} 
\newcommand{\fst}{{FeSe$_{0.5}$Te$_{0.5}$}\xspace}
\newcommand{\fesete}{{Fe(Se,Te)}\xspace}

\renewcommand{\Re}{\textrm{Re}}
\renewcommand{\Im}{\textrm{Im}}
\reversemarginpar

\usepackage{orcidlink}

\makeatletter
\long\def\@makefntext#1{\parindent 1em\noindent 
 \makebox[1em][l]{\footnotesize\rm$\m@th{\fnsymbol{footnote}}$}%
 \footnotesize\rm #1}
\def\@makefnmark{\hbox{${^\fnsymbol{footnote}}\m@th$}}
\def\@thefnmark{\fnsymbol{footnote}}
\makeatother

\usepackage[absolute]{textpos}
\textblockorigin{0in}{0in} 

\begin{document}
\title[{Flux flow and orbital upper critical field} in multiband \fst explored by microwave magnetotransport]{{Flux flow and orbital upper critical field} in multiband \fst explored by microwave magnetotransport}
\author{A. Magalotti$^{1,2}$~\orcidlink{0009-0004-3352-1977}, A. Alimenti$^{1,2}$~\orcidlink{0000-0002-4459-6147}, V. Braccini$^{3}$~\orcidlink{0000-0003-0073-367X}, P. Manfrinetti$^{4}$~\orcidlink{0000-0002-3346-5619}, E. Silva$^{1,2}$~\orcidlink{0000-0001-8633-4295}, K. Torokhtii$^{1}$~\orcidlink{0000-0002-3420-3864}, N. Pompeo$^{1,2}$~\orcidlink{0000-0003-4847-1234}}
\address{$^{1}$Department of Industrial, Electronic and Mechanical Engineering, Universit\`a Roma Tre, Via Vito Volterra 62, 00146 Roma, Italy}
\address{$^{2}$INFN, Roma Tre Section, 00146 Roma, Italy}
\address{$^{3}$CNR-SPIN, Corso Perrone 24, 16152 Genova, Italy}
\address{$^{4}$Department of Chemistry and Industrial Chemistry, Universit\`a Degli Studi di Genova, Via Dodecaneso 31, 16146 Genova, Italy}
\ead{nicola.pompeo@uniroma3.it}
\vspace{10pt}

\begin{textblock}{7}(1,10.5) 
\noindent\centering
This is the version of the article before peer review or editing, as submitted by an author to Superconductor Science and Technology. IOP Publishing Ltd is not responsible for any errors or omissions in this version of the manuscript or any version derived from it. The Version of Record is available online at \url{https://doi.org/10.1088/1361-6668/ae5151}.
\end{textblock}
\begin{abstract}
We measure the {flux flow} resistivity in \fst epitaxial films using a microwave dual-frequency technique (\qty{16}{\giga\hertz} and \qty{27}{\giga\hertz}), in the range \qty{5}{\kelvin}-$T_c$, in static magnetic fields up to \qty{1.2}{\tesla}.
By applying a temperature scaling procedure, we extract from {flux flow} measurements the temperature dependence of the orbital upper critical field, that shows features of multiband superconductors. The reduced orbital upper critical field is then fitted with a \text{two-band} model with strong intraband and weak interband coupling, as expected in 11 Fe-based systems. We derive the vortex viscosity and we estimate the \text{bands-averaged} vortex core quasiparticle reduced scattering time within the \text{Bardeen-Stephen} framework. Our data suggest that, for our epitaxial \fst films, the quasiparticle scattering rate values are at the upper edge for the dirty regime. Finally, tentative numerical values of the orbital upper critical field and coherence length are provided.
\end{abstract}

\ioptwocol

\section{Introduction}
Iron based superconductors (IBS) are well known to be multiband superconductors. Various pairing symmetries are proposed and have found experimental support, with more or less anisotropic s-wave gap with possible $++$ peculiarities \cite{hosono2015a}. The Fermi surfaces involves typically two or more bands, with both types of charge carriers (holes and electrons). Moreover, anisotropy is {an important} feature of IBS, albeit less prominent than in their cuprate high-$T_c$ cousins.

In the mixed state, vortex structure and dynamics is expected to be deeply impacted by this complex electronic structure, leading for example to fractional vortices, related to the distinct superfluids, which can be locked \cite{goryo2005} or lead to dissociate at high driving currents \cite{lin2013}. The multiband nature can also appear in terms of enhanced dissipation at low fields, whenever one of the bands is more strongly depressed by the applied magnetic field, similarly to what happens in MgB$_2$ \cite{goryo2005,sarti2005}. Indeed, the important parameter in this context, the slope $\alpha$ of the normalized {flux flow} resistivity $\rhoff/\rho_n$ vs the normalized applied magnetic field $B/B_{c2}$, ${\rhoff/\rho_n=\alpha B/B_{c2}}$, is expected and observed to be $\alpha >1$ ($\rho_n$ and $B_{c2}$ are the normal state resistivity and upper critical field, respectively) \cite{takahashi2012,okada2012,okada2013,okada2013a,okada2014}, in contrast with the well-known Bardeen-Stephen behaviour \cite{bardeen1965} or with more refined results based on Time-Dependent Ginzburg Landau theory \cite{schmid1966, LO1986}, although \fesete single crystals have exhibited contrasting behaviors \cite{okada2015}.

Further complexities arise when the quantization of the bound levels in the vortex cores is investigated, with the possible emergence of the superclean regime: in single-band superconductors the quantitative values of $\rhoff$ can be used as a marker of the superclean/clean regime \cite{golosovsky1996}. However in multiband superconductors the mixing of the bands contributions, with the addition of partially canceling effects \cite{ogawa2023a} whenever charge carriers of opposite signs are involved, leads to values for $\rhoff$ nominally pertaining to moderately clean regime even in {11-}systems like FeSe where the superclean limit is expected to be reached \cite{okada2021}.

The {flux flow} resistivity has been used in other superconductors to evaluate the Ginzburg-Landau coherence length $\xi$. However, IBS exhibit Pauli limited upper critical fields $\Bpau$ \cite{tarantini2011}, which flattens the temperature ($T$) dependence of $B_{c2}(T)$ by decreasing $T$. This property makes the determination of $\xi$ difficult, which instead is related to the so-called orbital upper critical field ${\Borb=\Phi_0/(2\pi\xi^2}$) \cite{tinkham1996} ($\Phi_0$ is the magnetic flux quantum). The evaluation of $\xi$ is made even more complex by the multiple superconducting order parameters coexisting in IBS.

The framework for the correct determination of the main superconducting parameters from $\rhoff$ is then rather {complex} in IBS, since all the peculiar elements of their electronic structure must be taken into account. In this work we study the microwave {flux flow} resistivity of \fst epitaxial films to (i) explore the degree of quantization of bound levels in the vortex cores exhibited by this IBS, (ii) shed light on the $\Borb$ temperature dependence, and (iii) assess whether the multiband superconductivity manifests in $\rhoff(B,T)$, and if differences exist with respect to observations in single crystals \cite{okada2015}.

The paper is structured as follows: \cref{sec:exp} introduces the microwave technique, the relevant models for the surface impedance and vortex motion resistivity, the samples preparation and properties, and the whole body of the surface impedance measurements. \cref{sec:results} deals with the analysis of the {flux flow} vortex motion parameters such as the {flux flow} resistivity and vortex viscosity and with the presentation and discussion of the main results. Conclusions are drawn in \cref{sec:conclusions}.

\section{Experiment and technique}\label{sec:exp}

\begin{figure}
\centering{\includegraphics[width=\columnwidth]{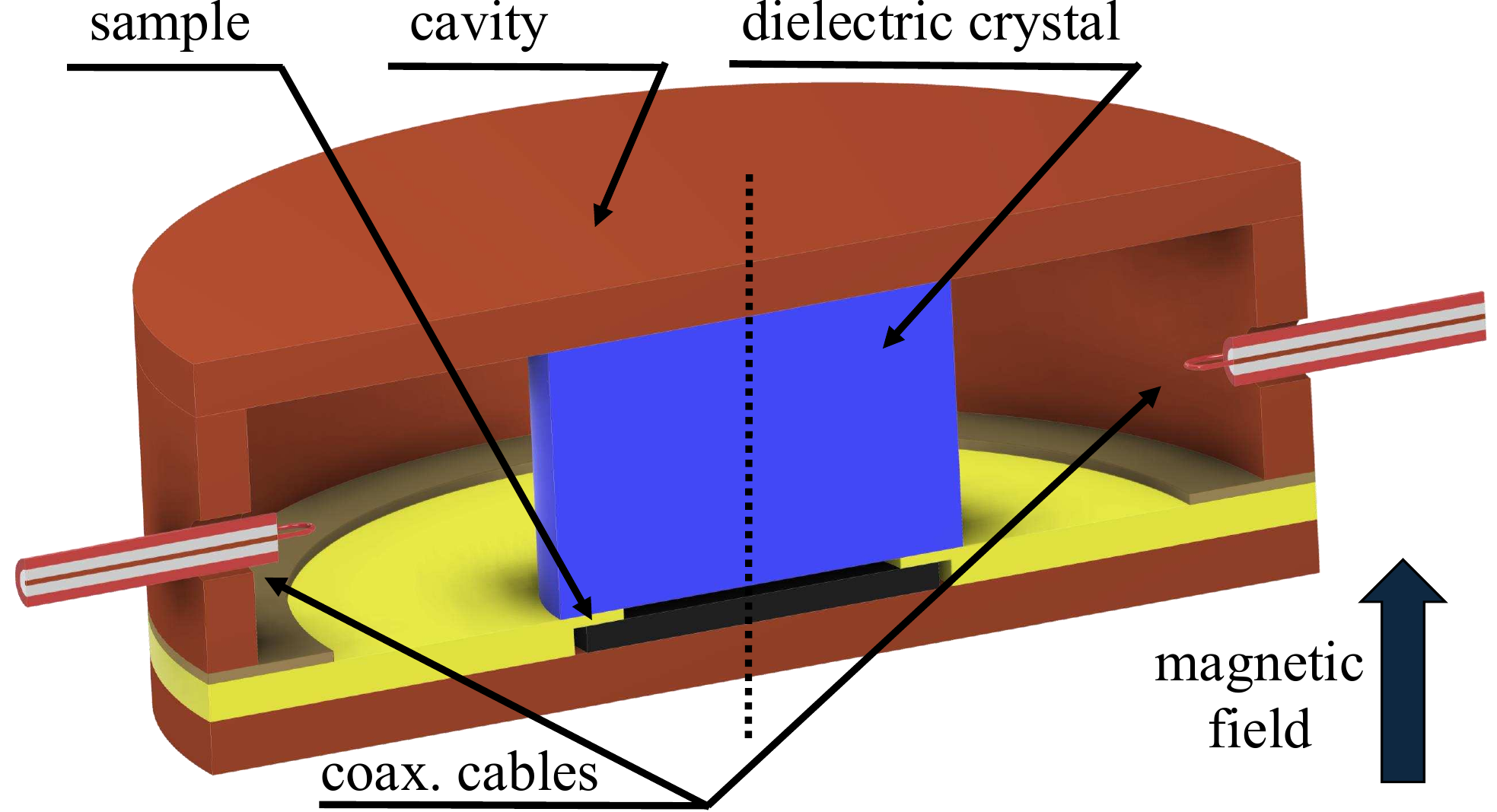}}
\caption{Dielectric loaded resonator used in the surface perturbation approach. The superconducting sample is depicted in black, the thin metal mask in yellow.}
\label{fig:resonator}
\end{figure}
\subsection{Microwave {t}echnique and {m}odels}
\label{sec:model}

We use a microwave technique to measure the surface impedance ${Z_s=E_{\parallel}/H_{\parallel}}$ of \fst thin films. Here, $E_{\parallel}$ and $H_{\parallel}$ are the microwave electric and magnetic field, respectively, parallel to the sample surface. $Z_s$ is the response function at microwave frequencies {\cite{pompeo2025}}. Measurements are performed by means of a cylindrical, dual mode dielectric-loaded resonator, with the \text{so-called} surface perturbation technique \cite{alimenti2019b}. A circular part of the sample surface is exposed to the resonator \text{e.m.} fields by means of a hole (diameter ${\approx\qty{6}{\milli\metre}}$) in a thin metal mask that partially covers it and holds it in a centered position (see \cref{fig:resonator}).

The two resonant modes used in the measurements are the TE$_{011}$ and TE$_{021}$, with resonant frequencies ${\nu_{0,1}=\qty{16.4}{\giga\hertz}}$ and ${\nu_{0,2}=\qty{26.6}{\giga\hertz}}$ respectively, thus well below the lowest gap energy exhibited by \fesete (${\sim\qty{2}{\milli\electronvolt} \simeq\qty{480}{\giga\hertz}}$ \cite{miao2012}) so that no pair breaking can be induced by the microwave probe apart from very close to $ T_c$. Both modes induce planar circular currents on the sample surface. The resonator is operated in transmission and its complex \text{two-port} scattering coefficients {$S_{jk}(\nu)$} are measured against frequency $\nu$ through a Vector Network Analyzer. With proper calibration and fitting procedures the resonant modes quality factors $Q_i$ and resonant frequencies $\nu_{0,i}$ (${i=1,2}$) are extracted \cite{torokhtii2020b}. From the \mbox{magnetic-field-induced} variations of $Q_i$ and $\nu_{0,i}$ at fixed temperature $T$, the corresponding variations of $\Delta Z_s(H; T)$ are determined as \cite{MicrowaveElectronics}:
\numparts
\begin{eqnarray}
\Delta Z_s(H)&=Z_s(H)-Z_s(0)\\
\Delta R_s(H)&=G_{i}\left(\frac{1}{Q_i(H)}-\frac{1}{Q_i(0)}\right)\\
\Delta X_s(H)&=-2G_{i}\left(\frac{\nu_{0,i}(H)-\nu_{0,i}(0)}{\nu_{0,i}(0)}\right)
\end{eqnarray}
\label{eq:Q_nu}
\endnumparts
where
\begin{equation}
\label{eq:Zs}
Z_s(H)=R_s(H)+\rmi X_s(H)
\end{equation}
and ${R_s=\Re(Z_s)}$ and ${X_s=\Im(Z_s)}$ are the surface resistance and reactance, respectively, and $G_{i}$ ${(i=1,2)}$ are the geometrical factors specific of each resonant mode. ${G_{1}={\qty{1334}{\ohm}}}$ and ${G_{2}={\qty{7455}{\ohm}}}$ were calculated through \text{e.m.} (electromagnetic) numerical simulation of the whole measurement cell, using the eigenmode solver of CST Studio Suite(R) 2025 software.

When measuring the temperature-induced variations $\Delta Z_s(T; H)$, an expression similar to Eq.~(1) holds, with the addition of a temperature dependent background term which needs to be properly taken into account \cite{alimenti2019b}.

The London penetration depth $\lambda_L$ is the lower limit for the penetration of the \text{e.m.} field. In \fesete compounds \cite{okada2015} \mbox{$\lambda_L(H\!=\!0;T\!=\!0)\!\simeq\!500\;$nm}, larger than the thickness $d$ of our films. Thus, the films are fully penetrated by the microwave field and the so-called thin film regime \cite{pompeo2025} takes place:
\begin{equation}
\label{eq:Zfilm}
Z_s\simeq\Zf=\frac{\tilde\rho}{d},
\end{equation}
{In this case, the surface impedance} is directly proportional to the sample complex resistivity $\tilde\rho$. In the mixed state the latter is given by the Coffey-Clem model \cite{coffey1991a}:
\begin{equation}
\label{eq:rhoc}
\tilde\rho=\frac{\rho_{vm}+\rmi/\sigma_2}{1+\rmi \sigma_1/\sigma_2}\simeq\rho_{vm}+\rmi\frac{1}{\sigma_2}=\rho_{vm}+\rmi{2\pi\nu\mu_0\lambda{_L}^2}
\end{equation}
The two fluid conductivity $\sigma_{2f}=\sigma_1-\rmi\sigma_2$ incorporates the quasiparticle contribution $\sigma_1$ and the superfluid contribution $\sigma_2=1/(2\pi\nu\mu_0\lambda{_L}^2)$. The approximate equality in \cref{eq:rhoc} holds far from the $B_{c2}(T)$ line, where ${\sigma_1\ll\sigma_2}$. The vortex motion resistivity $\rhovm$ is a complex, frequency dependent quantity which, on very general grounds \cite{Pompeo2008}, can be cast in the following form:
\begin{equation}
\label{eq:rhovm}
\rhovm(B)=\rho_{vm,1}+\rmi\rho_{vm,2}=\rhoff\frac{\chi+\rmi\nu /\nu_c}{1+\rmi \nu/\nu_c}
\end{equation}
where the vortex parameters $\rhoff$, $\nu_c$ and $\chi$ are the {flux flow} resistivity, the characteristic frequency and the thermal creep factor, respectively. With the measurement of $\rhovm(B; \nu)$ at the two operating frequencies $\nu_{0,i}$, \cref{eq:rhovm} yields four observables (real and imaginary parts at two frequencies) which can be used to extract the field dependent vortex parameters through analytical inversion of the equations, as done here following Ref.~\cite{pompeo2021}.
In this way the {flux flow} resistivity can be determined independently from pinning contributions, in contrast with what happens, for example, in dc magnetoresistivity measurements where {flux flow} and pinning effects are intermingled in the measured dc resistivity $\rho_{dc}$. It is worth noting that, since ${\rhovm(B=0)=0}$, when pair-breaking effects \cite{tinkham1996} are negligible so that $\lambda{_L}$ is independent from $B$, field-induced variations on $Z_s$ in the thin film regime directly yield $\rhovm(B)$:
\begin{equation}
\label{eq:DZfilm}
\Delta Z_s(B)=\frac{\rhovm(B)}{d},
\end{equation}
with $B=\mu_0 H$ in the London approximation.

\subsection{Samples preparation}\label{sec:sample}

\begin{table}
    \centering
    \renewcommand{\arraystretch}{1.3}
    \begin{tabular}{|c|c|c|c|c|}
        \hline
        Sample & {{$d\;$(\unit{nm})}} & $T_c\;(\unit{K})$ & $R_n\;(\unit{\ohm})$ & $\rho_n\;(\unit{\ohm \cdot \m})$ \\
        \hline
        FST\#1    & 240   & 18.0   & 13.0 & 3.1$\times10^{-6}$ \\ \hline
        FST\#2    & 400   & 18.6   & 7.5 & 3.0$\times10^{-6}$  \\ \hline
    \end{tabular}
    \caption{Studied samples with main parameters. The (onset) critical temperature $T_c$ is determined from the superconducting transition on $R_s$ as reported in \cref{fig:transition}. }
    \label{tab:sample_table}
\end{table}

\begin{figure}
\includegraphics[width=\columnwidth]{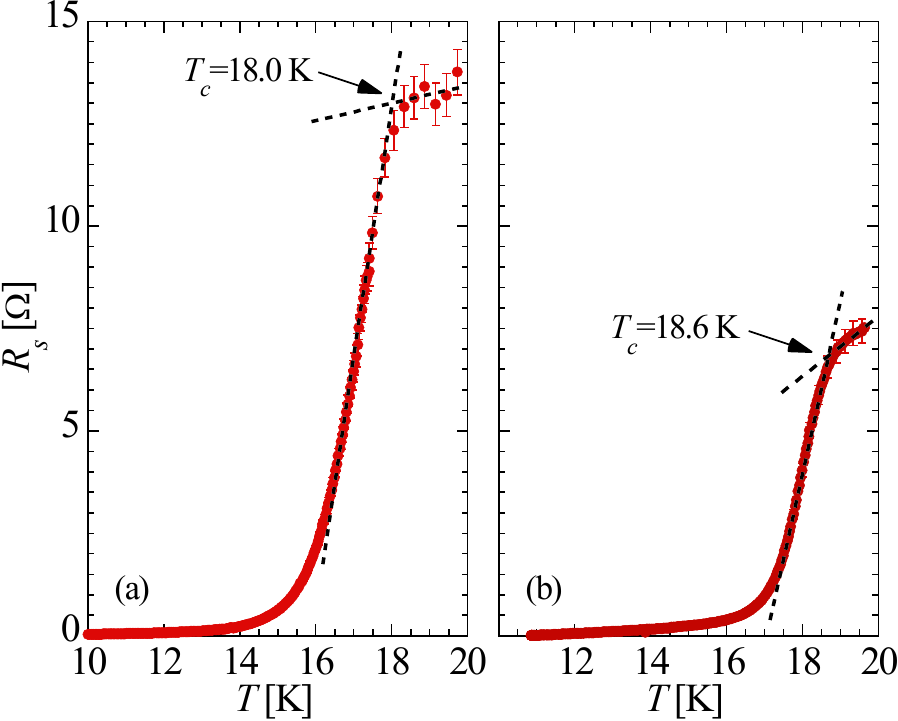}
\caption{Superconducting transitions $R_s$ vs $T$ at ${\nu_{0,2}=\qty{27}{\giga\hertz}}$ and at zero field. (a)~Sample FST\#1; (b) sample FST\#2.}
\label{fig:transition}
\end{figure}

The samples under study are \fst of nominal composition ${x=0.5}$ and were deposited by Pulsed Laser Deposition on CaF$_2$ substrates, $7\times7\;\unit{\milli\meter\squared}$ and \qty{0.5}{\milli\meter} thick. The deposition was carried out in a high vacuum PLD system equipped with a Nd:YAG laser at \qty{1024}{\nano\meter}, using a FeSe$_{0.5}$Te$_{0.5}$ target synthesized with a two-step method \cite{palenzona2012}. The laser parameters were optimized to obtain high quality epitaxial thin films: \qty{3}{\hertz} repetition rate, 2\;J$\cdot$cm$^{-2}$ laser fluency ($2\;\text{mm}^2$ spot size) and a \qty{5}{\centi\meter} \text{substrate-target} distance \cite{sylva2018}. During the deposition, the substrate was kept at \qty{350}{\degreeCelsius} at the base pressure of the system (1$\times$10$^{-8}\;$mbar). Superconducting transitions $R_s$ vs $T$ measured at ${\nu_{0,2}=\qty{27}{\giga\hertz}}$ and at zero field are reported in \cref{fig:transition}, whence the (onset) $T_c$, the normal state resistance $R_n$ and the normal state resistivity ${\rho_n=R_n d}$ as derived from microwave measurements, see \cref{eq:Zfilm}. The thickness has been calibrated against deposition time. Thus, a systematic uncertainty $\pm10\%$ on the thickness of each sample with respect to the nominal value is estimated. Sample data are reported in \cref{tab:sample_table}, where it can be seen that they exhibit normal state resistivity and $T_c$ in good agreement with other similar samples \cite{palenzona2012,pompeo2020a}. Nanostructural analysis, film quality assessment and XRD measurements, performed on very similar \fst thin films, grown in the same laboratory, with the same technique and substrate, are reported in \cite{scuderi2021} and \cite{Fracasso_2025}, respectively. Typical residual resistivity ratio in such samples are $RRR = 1.1\div1.5$.

\subsection{Surface impedance data}\label{sec:zsdata}

\begin{figure}
\includegraphics[width=\columnwidth]{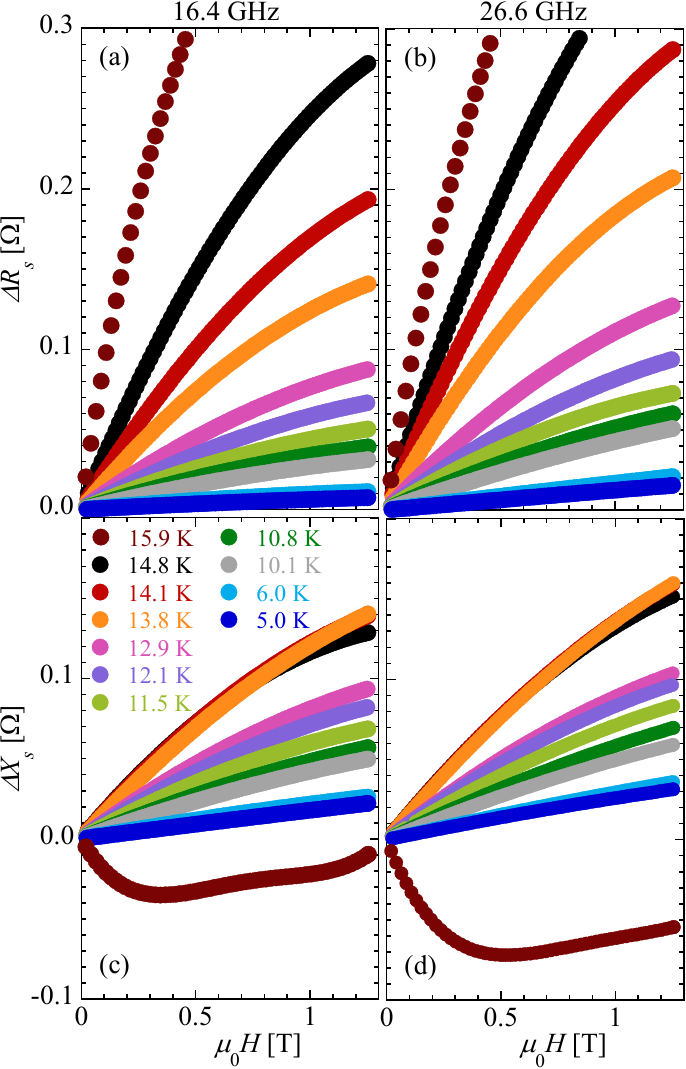}
\caption{Measured $\Delta Z_s(H)$ for TE$_{011}$ mode at \qty{16.4}{\giga\hertz} and TE$_{012}$ mode at \qty{26.6}{\giga\hertz}, at selected $T$. The measurements are taken with $\vec{H}\perp\;$sample surface. Data for sample FST\#1. Error bars are within symbol size.}
\label{fig:Zs_vs_field1}
\end{figure}

\begin{figure}
\includegraphics[width=\columnwidth]{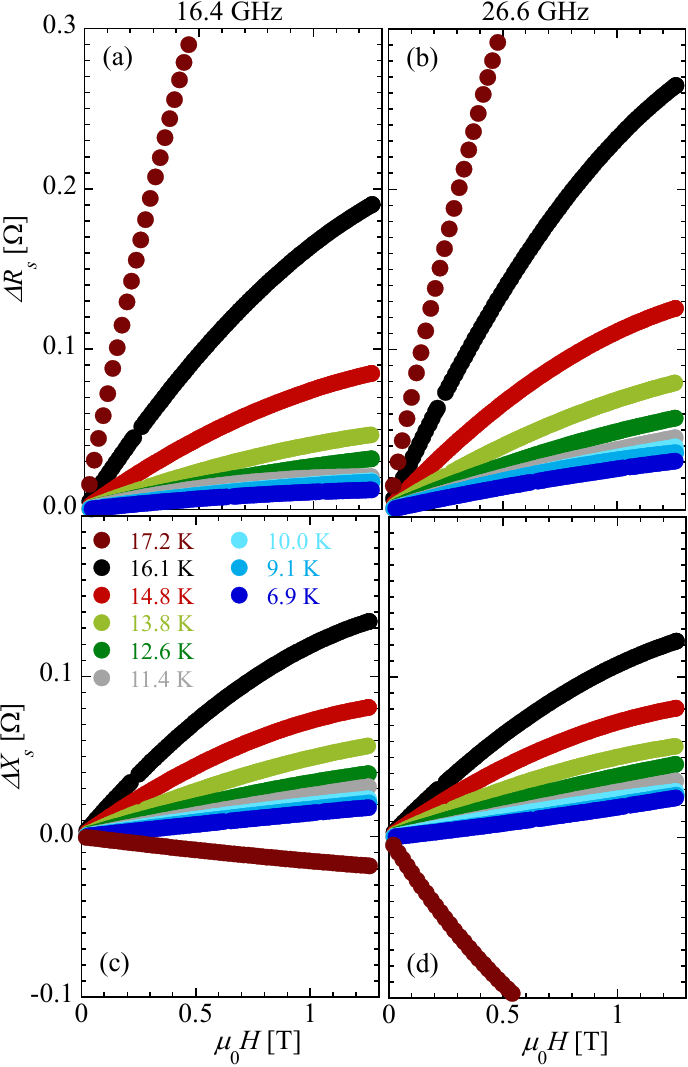}
\caption{Measured $\Delta Z_s(H)$ for TE$_{011}$ mode at \qty{16.4}{\giga\hertz} and TE$_{012}$ mode at \qty{26.6}{\giga\hertz}, at selected $T$. The measurements are taken with $\vec{H}\perp\;$sample surface. Data for sample FST\#2. Error bars are within symbol size.}
\label{fig:Zs_vs_field2}
\end{figure}

Field induced variations of the surface impedance, $\Delta Z_s(H)$, have been measured as a function of the static magnetic field in the range $H\in[0\;\text{T}, 1.2\;\text{T}]$, from \qty{5}{\kelvin} up to $T_c$. The field was applied perpendicularly to the sample surface, parallel to the crystallographic \text{$c$-axis} of the samples. Raw measurements performed at the two measuring frequencies ${\nu_{0,1}=\qty{16}{\giga\hertz}}$ and ${\nu_{0,2}=\qty{27}{\giga\hertz}}$ are fully reported in \cref{fig:Zs_vs_field1,fig:Zs_vs_field2}.

We obtain from \cref{eq:DZfilm} the vortex motion resistivity as ${\rho_{vm}=\Delta Z_s(H)\,d}$. Since the uncertainty on the thickness $d$ (10\%) is the same for all $T$, it does not contribute to the scaling procedure described in  \cref{sec:results}. Moreover, the systematic uncertainty on $d$ enters only as a common scale factor on $\rhoff$, which does not change the conclusions drawn in \cref{sec:qptau} for the analysis and thus it is not reported for this quantity. In this paper, the figures report error bars corresponding to standard uncertainties with confidence interval 1. The same analysis has been used to estimate the anisotropy \cite{pompeo2021b} and to thoroughly study the vortex pinning in films similar to those here presented \cite{pompeo2020b}. 

{In particular, it is here useful to recall that first, pinning shows a nearly negligible effect ${\sim\qty{2}{\kelvin}}$ below $T_c$, and second, for ${T\rightarrow T_c}$ the field-induced pairbreaking can give a significant contribution even at the modest dc fields here applied.} These factors{, which generally} complicate the analysis of the data{,} {can be here quite simply handled. In fact a clear signature of field-induced pair-breaking appears in the form of}
${\Delta X(H)<0}$ in the thin film limit {(see \cref{eq:Zfilm}): indeed, a ${\Im[\tilde\rho] \neq 0}$ at $H=0$ must become zero} at the superconducting transition. The precise field dependence is determined by the competition of pair breaking with all the possible details of pinning weakening. As recalled above, in our samples pinning nearly vanishes only very close to $T_c$ \cite{pompeo2020b}. Accordingly, our data show ${\Im[\Delta\tilde\rho (H)]\leq0}$ at the highest measured temperature only. In this case, \cref{eq:DZfilm} does not hold and the extraction of the vortex motion parameters requires a dedicated procedure, in which pairbreaking is included. On the basis of this consideration, the {flux flow} resistivity $\rhoff$ data at the highest measuring temperatures, ${T=\qty{15.9}{\kelvin}}$ in sample FST\#1 and ${T=\qty{17.2}{\kelvin}}$ in sample FST\#2, were calculated assuming a superposition of pure {flux flow} motion and superfluid pairbreaking effects \cite{tinkham1996}, see \cref{eq:rhoc}. We have carefully checked that the model holds and correctly predicts ${\Delta X(H)<0}$ while ${\Delta R(H)>0}$.

\section{Results}
\label{sec:results}
\begin{figure}
\centerline{\includegraphics[width=\columnwidth]{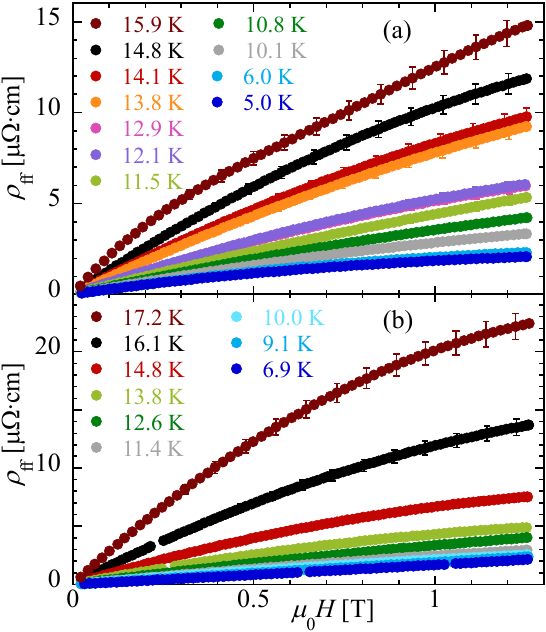}}
\caption{{flux flow} resistivity $\rhoff$ vs $H$, at selected $T$. (a)~{S}ample FST\#1; (b) sample FST\#2.}
\label{fig:rhoff_B}
\end{figure}

It is well known that at microwaves the {flux flow} resistivity $\rhoff$ can be directly measured, differently from dc measurements, independently of the fluxons pinning strength (see \cref{eq:rhovm}). The potential of microwave measurements to disentangle {flux flow} from pinning effects has been already exploited to describe and discuss the pinning strength in \fst thin films \cite{pompeo2020b}. In this Section we discuss $\rhoff$ and the physics that can be extracted. As described in \cref{sec:model}, using \cref{eq:rhovm,eq:DZfilm} and exploiting the combined measurements at two frequencies (see \cite{pompeo2021}), we extract $\rhoff$. The results are reported in \cref{fig:rhoff_B} for selected temperatures in both samples studied. We first note a qualitative fingerprint of nonstandard {flux flow}: the downward curvature of $\rhoff(B)$ (\cref{fig:rhoff_B}) is very similar to what observed in MgB$_2$ \cite{goryo2005,shibata2003,alimenti2023a,alimenti2025} and several IBS superconductors \cite{takahashi2012,okada2012,okada2014}. This is clearly only a qualitative hint, and we proceed with a more quantitative analysis.

\subsection{Quasiparticle scattering time in the vortex cores}\label{sec:qptau}
In the Bardeen-Stephen framework \cite{bardeen1965}, $\rhoff$ is directly linked to the quasiparticle scattering time $\tauQP$ in the vortex cores \cite{golosovsky1996} and thus it can be used to assess whether the superconductor falls in the clean or dirty limits. Let $\Delta E$ be the energy separation between the quantized bound levels in vortex cores \cite{caroli1964} and ${\delta E\approx \hbar/\tauQP}$ their width due to scattering processes. The ratio $\Delta E / \delta E$ is a measure of the vortex core cleanness: if ${\Delta E / \delta E\ll 1}$ (dirty regime), the spread of the quantized levels let them overlap yielding a continuous levels spectrum, whereas ${\Delta E / \delta E\gg1}$ yields the superclean regime. The in-between situation ${\Delta E / \delta E \sim 1}$ marks the so-called moderately clean regime. One has:
\begin{itemize}
 \item $\Delta E / \delta E \approx \omega_c \tauQP$ \cite{caroli1964,blatter1994}, being $\omega_c$ the cyclotron angular frequency \cite{golosovsky1996};
 \item $\omegatau{bands}=\eta_{eff}/(\pi\hbar n)$, with ${\eta_{eff}=\Phi_0 B /\rhoff}$ the effective vortex viscosity as measured in high frequency experiments with externally imposed (microwave) transport current density \cite{golosovsky1996} and $n$ the charge carrier density; $\langle \rangle_{\text{bands}}$ highlights that the quantity obtained from the vortex viscosity is averaged over the Fermi surfaces of all the bands possibly present.
\end{itemize}

\begin{figure}
\includegraphics[width=\columnwidth]{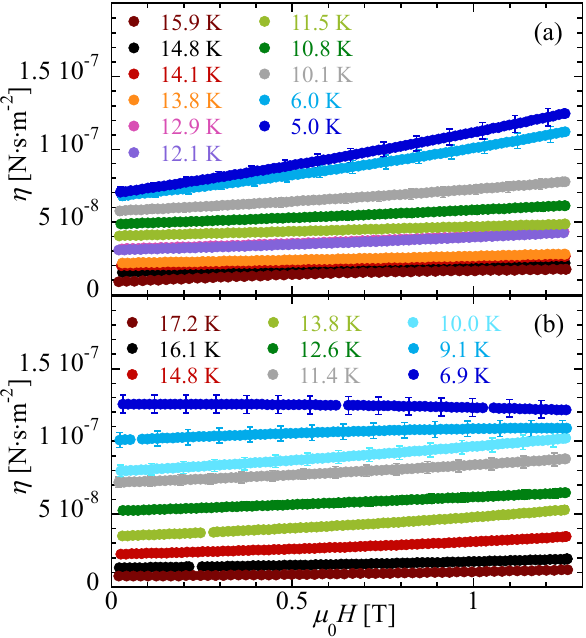}
\caption{Effective vortex viscosity vs $H$ at selected $T$. (a)~{S}ample FST\#1; (b) sample FST\#2.}
\label{fig:eta}
\end{figure}

Vortex viscosity data are reported in \cref{fig:eta}. It can be seen that the vortex viscosity has not a strong $B$ dependence (which corresponds to the weak downward curvature observed in $\rhoff(B)$, \cref{fig:rhoff_B}), indicating that in the present field range $B<\qty{1.2}{\tesla}$ the (effective) scattering rate is only weakly field dependent. This fact does not exclude that, reaching the upper critical field, scattering could increase, thus potentially providing increased interband scattering.

In \fesete one must consider hole and electron bands \cite{miao2012}. A mixture of the actual scattering times of the individual bands \cite{ogawa2023a}, $\omegatau{h}$ and $\omegatau{e}$, determines the \emph{effective} $\omegatau{bands}$.
The functional dependence of $\omegatau{bands}$ on $\omegatau{h,e}$ is rather intricate, as extensively discussed in Ref.~\cite{ogawa2023a} (to the best of our knowledge, the only work addressing extensively the present question). However, some considerations can be made in simple cases. For two bands in the dirty limit, ${\omegatau{h,e} \ll 1}$ (hydrodynamic regime), the effective viscosity ${\eta_{eff}=\eta_h+\eta_e}$ \cite{ogawa2023a}. For identical bands, one would have $\eta_{eff}$ twice the value of one of the single-band $\eta$. For very different $\eta_{h,e}$, $\eta_{eff}$ would approximately correspond to the largest. This qualitative picture remains approximately valid also when the bands are not deep in the dirty limit, with complications given by Hall terms giving partial cancellations \cite{ogawa2023a}.

To estimate $\omegatau{bands}$, the hole and electron carrier densities, $n_h$ and $n_e$, respectively, are needed. Estimates for $n_{h,e}$ have been reported in literature \cite{nabeshima2020,Nabeshima2018,otsuka2019,zhang2025,terashima2016}, and  depend on the composition of \fst. For samples of composition closer to our samples, $n_h=3.5\times10^{26}\;$m$^{-3}$ and $n_e=2.5\times10^{26}\;$m$^{-3}$ were reported \cite{nabeshima2020}. It is an important remark that $n_h$ and $n_e$ do not differ much in this case, so that important differences in $\eta_{h,e}$, if any, are exclusively due to differences in $\omegatau{h,e}$. We now assume $n=(n_h+n_e)/2$ without a large loss of generality since $n_h$ and $n_e$ are similar. Should the bands have very similar $\eta_{h,e}$, this choice would bring an overestimate for the very similar  $\omegatau{h} \simeq  \omegatau{e}$ by a factor $\sim2$. Should one band have a much larger $\eta$, this choice would introduce an uncertainty of $\sim 20\%$ on the corresponding estimate for  $\omegatau{h,e}$ for that band, since we ignore which concentration pertains to that band.

\begin{figure}
\includegraphics[width=\columnwidth]{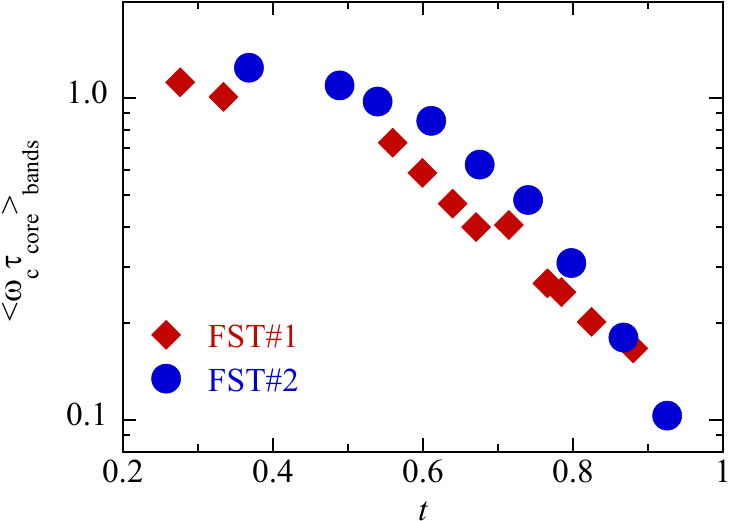}
\caption{Effective normalized vortex core scattering time, $\omegatau{bands}$ vs {$t$, the reduced temperature ${t = T/T_c}$,} in \fst for samples FST\#1 and FST\#2. The data {values are at the upper edge for the dirty regime}. Data calculated for $\mu_0H=$\qty{1.0}{\tesla}.}
\label{fig:omegatau}
\end{figure}

The obtained $\omegatau{bands}$ vs $T$ at ${B=\qty{1.0}{\tesla}}$ are reported in \cref{fig:omegatau} for the two samples.
Within the limitations described above, the values for $\omegatau{bands}$ place our samples at the upper limit of the dirty regime at low $T$. The values ${\omegatau{bands}\simeq{0.1\div}1}$ are not far from ${\omega_c \tauQP={1.0\pm0.5}}$ determined in FeSe \cite{okada2021}. One should note that the estimates for $\omegatau{bands}$ reported in Fig.~\ref{fig:omegatau} are obtained with the choice of $n=(n_h+n_e)/2$. According to the previous discussion, these numbers represent an estimate for the cleaner band quasiparticle scattering time when $\omegatau{bands}\sim 1$, while at low values we cannot exclude that they are an overestimate by $\sim 2$ for $\omega_c\tauQP$ in dirtier bands.

\subsection{The {flux flow} resistivity scaling: orbital upper critical field T-dependence}\label{sec:rhoff}

\begin{figure}
\centerline{\includegraphics[width=\columnwidth]{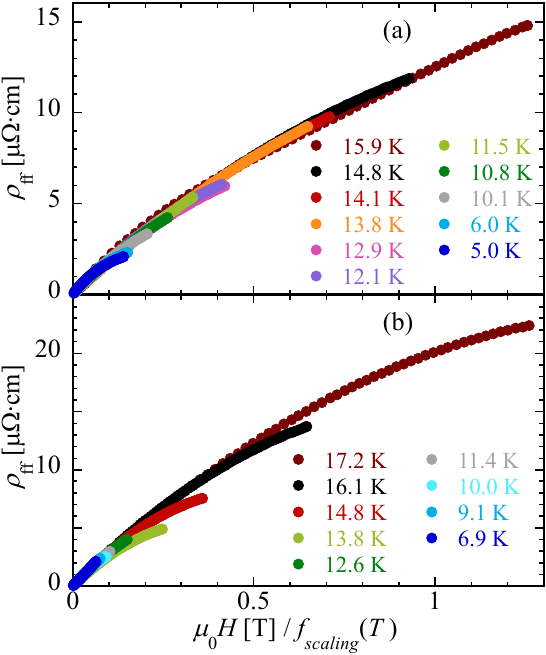}}
\caption{Scaling of the {flux flow} resistivity $\rhoff$ vs $H$.
(a)~{S}ample FST\#1; (b) sample FST\#2. A suitable temperature-dependent scaling factor $f_{scaling}(T)$ allows to satisfactorily scale all measurements at low fields.}
\label{fig:rhoff_Bscaling}
\end{figure}

The main fingerprint of multiband superconductivity can be found in the temperature dependence of the upper critical field. As remarkably demonstrated by Y. B. Kim, $et\;al.$ \cite{kim1965flux}, at low magnetic fields, $i.e.$ when ${B\ll \Borb\;(\text{or }\Bpau)}$ $\rhoff$ is inversely proportional to $\Borb$ only, even in strongly Pauli limited superconductors. In such regime, $\rhoff$ is defined with the well-known expression:
\begin{equation}
\label{eq:alpha}
\rhoff=\rho_n\alpha \frac{B}{\Borb} = \rho_n\alpha B \frac{2\pi\xi^2}{\Phi_0},
\end{equation}
where $\alpha$ is a parameter linked to the microscopic details of the current patterns in and around the moving vortices, the quasi-particle (QP) scattering times and density, the actual energy levels available in the vortex cores, the anisotropy of the Fermi Surface (FS) and the single/multiband nature of the superconductor \cite{schmid1966,LO1986,golosovsky1996,caroli1964,dorsey1992,kopnin1997}. $\alpha$ is then an essential parameter to study the microscopic nature of a superconductor. To compute $\alpha$ from $\rhoff$, an estimate of $\xi$ (or $\Borb$) entering \cref{eq:alpha} is needed \cite{kopnin2001a}. A direct measure of $\Borb$ in IBS is difficult: in fact, it is well known that IBS, and \fesete in particular, can have a strongly Pauli limited ${B_{c2} = \Bpau\ll\Borb}$ \cite{tarantini2011,grimaldi2019a}, in particular at low $T$, so that $B_{c2}$ typically steeply rises near $T_c$ and then flattens as $T\rightarrow0$.

\begin{figure}
\centerline{\includegraphics[width=\columnwidth]{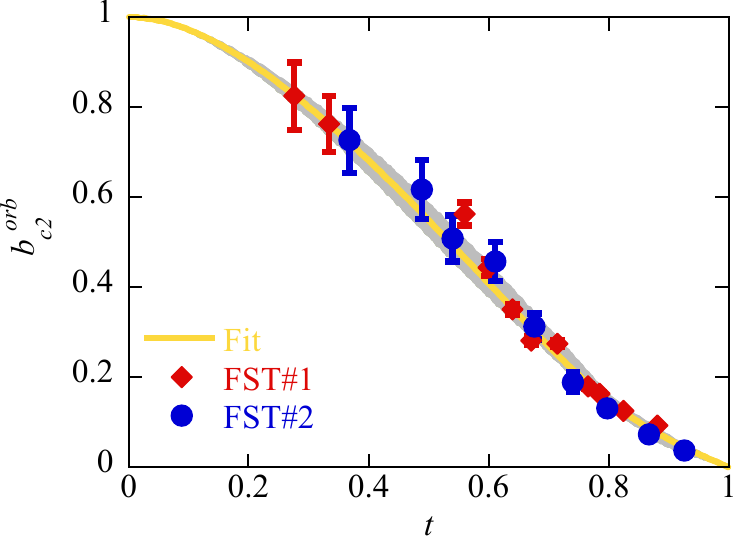}}
\caption{Estimated reduced critical field $b_{c2}^{orb}(t)$ vs reduced temperature $t$ for both samples. The continuous line is the Gurevich model (Eq.~(\ref{eq:2band})) fit for the two samples. The fitting parameters are ${\lambda_{11}=0.65\pm0.01}$, ${\lambda_{22}=0.61\pm0.01}$, ${\lambda_{12}=\lambda_{21}=0.06\pm0.04}$ and ${\zeta=0.25\pm0.05}$. The shaded area shows the fit variation when the parameters are changed within the given uncertainties.}
\label{fig:Bc2orb}
\end{figure}

We then exploit the magnetic-field-scaling properties of $\rhoff$: {if a} negligible $T$ dependence of $\alpha$ and $\rho_n$ {is assumed},\footnote{{The weak reported linear $T$ dependence of $\rho_n$ \cite{Fracasso_2025,braccini_private} corresponds to a maximum $10\%$ variation if linearly extrapolated below $T_c$. We have estimated a worst-case scenario where $f_{scaling}(T)$ would be impacted by less than $4\%$, and variations of the fit parameters discussed in the following paragraph would be contained within their uncertainty bars. The present data do not allow to determine separately $\alpha(T)$ and $B_{c2}(T)$. Nevertheless, the few existing data for various IBS compounds \cite{okada2012,okada2013,okada2013a}, show zero or very weak $T$ variations.}}
its only $T$ dependence comes from $\Borb$ so that ${\rhoff(B, T)= \rhoff(B/\Borb(T))}$. Hence, a scaling function $f_{scaling}(T)$ can be determined so that the various experimental curves $\rhoff(B/f_{scaling}(T))$ superimpose. 
{The $\alpha(T)$ dependence can be safely ignored, since no sign of band suppression is observed at low $B$. Conversely, $\rho_n(T)$ has been experimentally calculated from \cite{Fracasso_2025,braccini_private} and removed from the scaling.} Such scaling is satisfactory, in particular at low fields, as reported in \cref{fig:rhoff_Bscaling} for both samples. The so-obtained $f_{scaling}(T)$ directly yields the temperature dependence of $\Borb(T)$ apart from an overall scale factor only. Once $f_{scaling}(T)$ is obtained, since it reflects the temperature dependence of $B_{c2}^{orb}(T)$, it can be converted in the normalized orbital upper critical field ${b^{orb}_{c2}(T) = B_{c2}^{orb}(T)/B_{c2}^{orb}(0)}$ by knowing that ${b^{orb}_{c2}(0) = 1}$. The result is reported in \cref{fig:Bc2orb}. As it can be seen, $b_{c2}^{orb}(T)$ agrees very well for both samples, with a temperature dependence which {exhibits an evident change of curvature at ${t = T/T_c\approx0.8}$, a typical signature of multiband superconductivity \cite{zehetmayer2013}. 
Another hallmark of multicomponent superconductivity is the temperature dependence of the anisotropies of different physical quantities like the London penetration depth or the critical fields \cite{zehetmayer2013}. Indeed, \fesete films similar to those studied here have shown an evident temperature dependent $B_{c2}$ anisotropy factor $\gamma_H(T)$, as reported in \cite{grimaldi2019a}.} 
{Hence, we look for a quantitative agreement of the observed $b^{orb}_{c2}(t)$ with a two-band framework.}
By exploiting a field scaling approach we assume that the single field scale involved is given by an overall $B_{c2}$. This approach is consistent with theories where finite interband coupling, albeit possibly small, gives rise to a $B_{c2}$ that cannot be separated as a simple combination of single-band upper critical fields. We then compare the $T$ dependence of our field scale to the {multiband theory developed by A. Gurevich \cite{gurevich2003} for} two-band superconductors with ${\omega_c\tau<1}$ in an extended temperature range. {Since $\Borb$ only is expected to be involved here, we do not include Pauli limiting effects into the fit, as instead done in \cite{Gurevich_2011}.} Different approaches \cite{goryo2005} have been used for clean superconductors at low $T$ \cite{okada2020} or for cases where a much weaker band could be driven in the quasiparticle state with the magnetic field, e.g. in MgB$_2$ \cite{alimenti2025}. Our \fst samples are not in the clean regime (\cref{sec:qptau}). In this case, the normalized orbital critical field can be obtained by solving the transcendental equation:
\begin{eqnarray}
\label{eq:2band}
&a_0 [\ln t + U(\gamma)] [\ln t + U(\zeta \gamma)] \; + \nonumber \\
&a_{{1}} [\ln t + U(\gamma)] + a_{{2}} [\ln t + U(\zeta \gamma)] = 0
\end{eqnarray}
where
\begin{itemize}
    \item $t=T/T_c$ is the reduced temperature;
    \item ${\zeta = D_2 / D_1}$ is the ratio of the {intraband} diffusivities;
    \item {$\gamma = \Borb{(t)} \hbar D_1 /(2\phi_0 k_B t {T_c})$ is an adimensional quantity for band 1 (whereas $\zeta\gamma$ is the respective quantity for band 2)}, with $\hbar$ the reduced Plank constant and $k_B$ the Boltzmann constant;
    \item $U(x) = \Psi(1/2 + x) - \Psi(1/2)$, with $\psi$ the Digamma function;
    \item $a_{0,1,2}$ are coefficients determined by the BCS coupling constants $\lambda_{ij}$ (where $i, j = 1,2$):
    \begin{itemize}
        \item $\lambda^\star = \sqrt{(\lambda_{11} - \lambda_{22})^2 + 4\lambda_{12}\lambda_{21}}$
        \item $a_0 = 2(\lambda_{11}\lambda_{22} - \lambda_{12}\lambda_{21}) / \lambda^\star$
        \item $a_1 = 1 + (\lambda_{11} - \lambda_{22}) / \lambda^\star$
        \item $a_2 = 1 - (\lambda_{11} - \lambda_{22}) / \lambda^\star$
        \end{itemize}
\end{itemize}
For \fesete, strong intraband {($\lambda_{11}$, $\lambda_{22}$)} and weak interband coupling {($\lambda_{12}$, $\lambda_{21}$)} are expected \cite{komendova2012,khasanov2010}. To reduce the number of free parameters, we assume $\lambda_{12}\lambda_{21} = \lambda_{12}^2$ as done in \cite{pan2024novel}. Thus, only four unknown parameters need to be determined, respectively  $\lambda_{11}$, $\lambda_{22}$, $\lambda_{12}$ and $\zeta$. In \cref{fig:Bc2orb}, the best fit is reported in solid line, with the values of the fit parameters reported in the caption. {As a further check of the robustness of the present approach, we have extensively checked that no standard single band $B_{c2}(t)$ fit, like WHH theory \cite{werthamer1966} or Ginzburg-Landau expansions (as done in \cite{klein2010thermodynamic}), were able to reproduce the temperature dependence of $b_{c2}^{orb}(t)$.} The fit uncertainty can be evaluated by examination of the shaded area, that represents the envelope of the fitting curves with the parameters in the reported range. The obtained parameters are in excellent agreement with those reported in \cite{pan2024novel,cho2011precision}, for \fesete samples with similar composition. We stress that $b_{c2}^{orb}(T)$ measurements at low temperatures are usually impossible to perform with conventional transport measurements, since $\Bpau$ ultimately limits $B_{c2}$. Therefore, the present analysis allows to unveil, in a complementary fashion, aspects of the vortex physics not explored  with dc transport measurements.

\begin{figure}
\centerline{\includegraphics[width=\columnwidth]{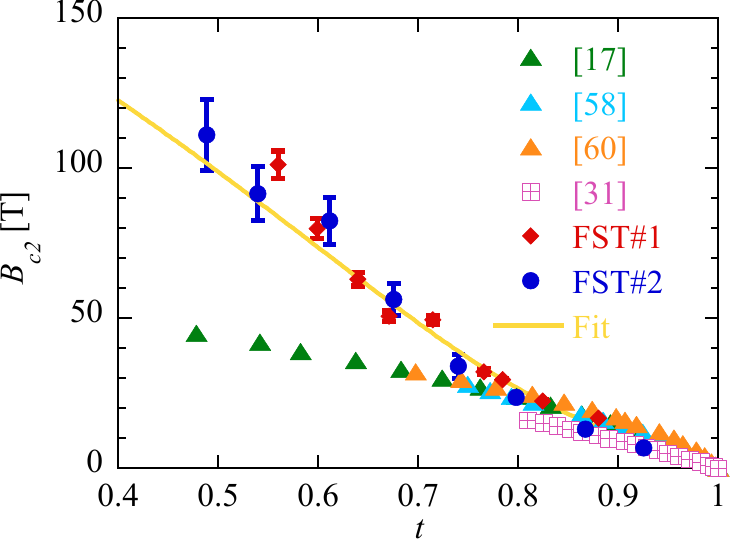}}
\caption{{Upper critical field vs ${t=T/T_c}$, comparison between the data presented in this paper, samples FST\#1 and FST\#2, and literature results. The low-field data are not affected by the Pauli limit, and thus our data agree with literature data. At higher fields, our procedure unveils the orbital upper critical field, while direct measurements report the Pauli-limited values (see text for details).}}
\label{fig:Bc2comp}
\end{figure}

It is possible to give tentative evaluations of the absolute values of $\Borb$, and thus $\alpha$ and $\xi$. Since at ${t>0.85}$ only orbital limit is expected to manifest \cite{Gurevich_2011}, we can match our $b_{c2}^{orb}(t)$ to $B_{c2}$ data for ${t>0.85}$ obtained from dc transport measurements performed on very similar \fst epitaxial thin films grown in the same laboratory,\footnote{{To determine $B_{c2,dc}$, we use here both the criterion for the electrical resistance ${R=0.5\cdot R_n}$ and ${\text{max}(dR/dT)}$ as a compromise between a higher threshold, likely to be affected by superconducting fluctuations, and a lower threshold, certainly affected by pinning and the irreversibility field.}}
with the same technique and substrate \cite{Fracasso_2025}. We obtain {$\Borb(0)=\qty{(180\pm10)}{\tesla}$} (the uncertainty is mainly given by the scattering of the two data sets). {The $\Borb$ values are then compared with other independent experimental data calculated from dc resistivity and specific heat measurements on \fst samples \cite{tarantini2011,klein2010thermodynamic,putti2010new}, together with the data from \cite{Fracasso_2025} mentioned above. The comparison is reported in \cref{fig:Bc2comp}, where the $B_{c2}$ values are plotted in absolute values. The deviation of the literature data for $B_{c2}$ with respect to the $\Borb$ here calculated, as soon as $T < 0.8\,T_c$, is evident.}

The estimates of $B_{c2}^{orb}$ yield ${0.8<\alpha<0.9}$, close to ${\alpha\sim0.66\div0.78}$ recently found by T. Okada, $et\;al.$ in {FeSe$_{0.4}$Te$_{0.6}$} single crystals \cite{okada2015}, where an explanation was given in terms of important scattering by magnetic impurities, presumably due to an excess of Fe. Although the field dependence of $\rhoff$ shows indeed the steep raise at low fields observed in other two-band superconductors, such as MgB$_2$ \cite{sarti2005,alimenti2025} and several IBS \cite{okada2015}, the obtained $\alpha < 1$ indicates that this value alone cannot be taken as a proof of presence or absence of multiband superconductivity.

Finally, we estimate the zero-temperature coherence length ${\qty{1.3}{\nano\meter}<\xi<\qty{1.4}{\nano\meter}}$. This result is in very close agreement with other estimates for \fesete, obtained also with techniques different from electrical transport \cite{tarantini2011,Leiner2014,Singh2013}.

\section{Conclusions}
\label{sec:conclusions}
Using a dual-frequency microwave technique, we successfully isolated the {flux flow} resistivity ($\rhoff$) from pinning influence in epitaxial \fst thin films grown on CaF$_2$. From $\rhoff$, we calculated the vortex viscosity and explored the \fst vortex core quasiparticle excitations, revealing averaged quasiparticle scattering times values pertaining to the upper edge for the dirty regime. The temperature-dependent scaling analysis of $\rhoff$ allowed to estimate the temperature dependence of the orbital upper critical field $\Borb(T)$, a parameter otherwise obscured by Pauli limit.~Clear signature of multiband superconductivity emerged in $\Borb(T)$, which were quantitatively described by an appropriate two-band model with strong intraband and weak interband coupling. Consequently, estimates of the coherence length were given. Our findings demonstrate that low magnetic field microwave measurements are a powerful tool to explore magnetic properties of Pauli limited or multiband superconductors, which are otherwise inaccessible via conventional dc transport measurements.

\ack
This work was partially supported by MIUR-PRIN Project ``HIBiSCUS'' Grant 201785KWLE, MIUR-PRIN Project  IronMOON Grant No. 2022BPJL2L, and INFN-CSN5 ``SuperMAD''.

\section*{References}
\providecommand{\newblock}{}

\end{document}